\documentstyle[prd,aps,epsfig]{revtex}

\newcommand{\be}{\begin{equation}}
\newcommand{\ee}{\end{equation}}
\newcommand{\ba}{\begin{eqnarray}}
\newcommand{\ea}{\end{eqnarray}}

\begin{document}
\draft
\title{Electron-electron Bound States in Parity-Preserving QED$_{3}$}
\author{H. Belich Jr.$^{a,b}$, O.M. Del Cima$^{a}$, M.M. Ferreira Jr.$^{a,c}$ and
J.A. Helay\"{e}l-Neto$^{a,b}$}
\address{$^{a}${\it Grupo de F\'{i}sica Te\'{o}rica Jos\'{e} Leite Lopes }\\
Petr\'{o}polis - RJ - \ Brazil.\\
$^{b}${\it Centro Brasileiro de Pesquisas F\'{i}sicas (CBPF)},\\
Coordena\c{c}\~{a}o de Teoria de Campos e Part\'{i}culas (CCP), \\
Rua Dr. Xavier Sigaud, 150 - Rio de Janeiro - RJ 22290-180 - Brazil.\\
$^{c}${\it Universidade Federal do Maranh\~{a}o (UFMA)}, \\
Departamento de F\'{i}sica, Campus Universit\'{a}rio do Bacanga, S\~{a}o\\
Luiz - MA, 65085-580 - Brazil. }
\maketitle

\begin{abstract}
By considering the Higgs mechanism in the framework of a parity-preserving
Planar Quantum Electrodynamics, one shows that an attractive
electron-electron interaction may come out. The $e^{-}-e^{-}$ interaction
potential emerges as the non-relativistic limit of the M{\o }ller scattering
amplitude and it may result attractive with a suitable choice of parameters.
Numerical values of the $e^{-}-e^{-}$ binding energy are obtained by solving
the two-dimensional Schr\"{o}dinger equation. The existence of bound states
is to be viewed as an indicative that this model may be adopted to address
the pairing mechanism in some systems endowed with parity-preservation.
\end{abstract}


\section{Introduction}

In the latest 10 years, Planar Quantum Electrodynamics - QED$_{3}-$ has
shown to be an appropriate theoretical framework for discussing the
low-energy limit of some Condensed Matter systems. Recent applications of
this theory to underdoped high-T$_{c}$ superconductors \cite{QED3} has again
caught attention for its theoretical possibilities. The history of the
relation between QED$_{3}$ and superconductivity goes back to the final 80%
\'{}%
s, when the anyonic model was established by the works of Laughlin \cite
{Laughlin}, and others \cite{Others}. Despite its initial success, it was
afterwards demonstrated that anyonic model supports the superconducting
phase only at zero temperature \cite{Lykken}. An alternative approach, also
based on the QED$_{3}$ framework, began to be adopted by Kogan \cite{Kogan}
to explain the formation of electron-electron bound states. Into the domain
of the QED$_{3},$ there exits the necessity of yielding a mass to the gauge
field in order to circumvent the appearance of a confining potential
associated to the long-range Coulombian interaction. The
Maxwell-Chern-Simons (MCS)\ term \cite{DJ} is then introduced as the
generator of (topological) mass for the photon, implying a screening on the
Coulomb interaction. This MCS-QED$_{3}$ model was used by some authors \cite
{Kogan}, \cite{Girotti} as basic tool for evaluation of the M\"{o}ller
scattering amplitude at tree-level, whose Fourier transform (in the Born
approximation) yields the $e^{-}e^{-}$ interaction potential. In a general
way, these works furnish the same result:\ the $e^{-}e^{-}$ interaction
comes out attractive when the topological mass $\left( \vartheta \right) $
surpasses the electronic mass $\left( m_{e}\right) $, that is, $\vartheta
>m_{e}$. This condition prevents the applicability of the MCS model to
solid-state systems, since the existence of a physical excitation with so
large energy in the domain of a Condensed Matter system is entirely
unlikely. It is possible to argue that the introduction of the Higgs
mechanism in the context of the MCS electrodynamics \cite{Int-Journal}, \cite
{Tese} brings out a negative contribution to the scattering potential that
will make feasible a global attractive potential regardless the condition $%
\vartheta >m_{e}$.

Some preliminary elucidations on the general character of this paper are
noteworthy. In spite of dealing with electron-electron condensation (a
mechanism that takes place in the realm Condensed Matter Physics), the
approach adopted here does not follow the usual procedure of Solid State
Physics, where one usually extracts information from the crystal lattice
about the interactions of the system. Indeed, our intention is not to follow
a solid-state approach, but to show instead that, starting from a
quantum-field theoretical framework, one may obtain (in the non-relativistic
limit) an attractive interaction that could be able to favour the electronic
pairing. In this way, we really develop a typical field-theoretical
approach, namely: one starts with a relativistic parity-preserving QED$_{3}$
Lagrangian (without MCS term) \cite{Tese}, \cite{N.Cimento}, \cite{Delcima}
containing fermions, scalar and vector bosons, and endowed with spontaneous
symmetry breaking (SSB). After the SSB, a Higgs boson and a massive photon
appear in the spectrum in the broken phase. These two particles mediate the
M\"{o}ller scattering, whose amplitude leads to a Bessel-$K_{o}$ interaction
potential, that can be attractive (independent of the electron polarization)
whenever the negative contribution stemming from the Higgs scalar
interchange dominates over the repulsive gauge interaction. It is also
important to stress that the explicit use of the Higgs-scalar interaction
(generated from the SSB) as an active mediator in the M\"{o}ller scattering
is the key factor for the attainment of an attractive potential. Relying on
the nonrelativistic approximation, the $K_{o}-$ potential is inserted into
the Schr\"{o}dinger equation. Its numerical solution provides us with values
of the $e^{-}e^{-}$ pairing energy,\ which are exhibited in Table \ref
{table1}.

This paper is organized as follows. In Sec. II, we present the
parity-preserving QED$_{3}$ model, explore some of its properties, and
realize the SSB. We then follow a field-theoretical development that leads
to the $e^{-}e^{-}$ interaction potential (derived in the non-relativistic
limit). In Sec. III, we accomplish a numerical procedure in order to obtain
numerical values for the binding energy of $e^{-}e^{-}$ pairs. In this
sense, we implement the $K_{o}$-potential into the Schr\"{o}dinger equation
and solve it. The numerical data are then displayed in Table I. In Sec. IV,
we present our Final Remarks.

\section{Parity-Preserving QED$_{3}$ With Spontaneous Symmetry Breaking}

We start with a parity-preserving QED$_{3}$ action (with SSB) \cite
{N.Cimento},\cite{Delcima}, built up by two polarized fermionic fields ($%
\psi _{+},\psi _{-}$) \cite{N.Cimento},\cite{Binegar}, a gauge potential $%
\left( A_{\mu }\right) $ and a complex scalar field $\left( \varphi \right) $%
:\ 
\begin{equation}
S=\int {d^{3}x}\biggl\{-{\frac{1}{4}}F^{\mu \nu }F_{\mu \nu }+{\overline{%
\psi }_{+}(i}{\rlap{\hbox{$\mskip 4.5 mu /$}}D}-m_{e}){\psi }_{+}+{\overline{%
\psi }_{-}(i}{\rlap{\hbox{$\mskip 4.5 mu /$}}D+}m_{e}){\psi }_{-}-y(%
\overline{\psi }_{+}\psi _{+}-\overline{\psi }_{-}\psi _{-})\varphi ^{\ast
}\varphi +D^{\mu }\varphi ^{\ast }D_{\mu }\varphi -V(\varphi ^{\ast }\varphi
)\biggr\},  \label{action1}
\end{equation}
with the scalar self-interaction potential, $V$, responsible for the SSB,
taken as: $V(\varphi ^{\ast }\varphi )=\mu ^{2}\varphi ^{\ast }\varphi +%
\frac{\zeta }{2}(\varphi ^{\ast }\varphi )^{2}+\frac{\lambda }{3}(\varphi
^{\ast }\varphi )^{3}$.$\ $ This sixth-order potential is the most general
one renormalizable in $1+2$ dimensions \cite{Delcima}. The mass dimensions
of the parameters $\mu $, $\zeta $, $\lambda $, $y$ are respectively ${1}$, $%
{1}$, ${0}$, ${0}${,} and the covariant derivatives, ${%
\rlap{\hbox{$\mskip 4.5
mu /$}}D}\psi _{\pm }\equiv (\rlap{\hbox{$\mskip 1
mu /$}}\partial +ie_{3}\rlap{\hbox{$\mskip 3 mu /$}}A)\psi _{\pm },$ $D_{\mu
}\varphi \equiv (\partial _{\mu }+ie_{3}A_{\mu })\varphi $, state the
minimal coupling between $\psi _{\pm },A_{\mu },$ and $\varphi $. It is
important to point out that the $U(1)-$symmetry coupling constant in $(1+2)$%
-dimensions, $e_{_{3}}$, has dimension of (mass)$^{\frac{1}{2}}$,\ a
relevant fact that must to be properly considered at the moment one needs to
attribute numerical values to it. We are interested only on a stable vacuum,
for which the following conditions on the potential parameters have to be
fulfilled: $\lambda >0~,~\zeta <0,$ $\mu ^{2}\leq {\frac{3}{16}}{\frac{\zeta
^{2}}{\lambda }.}$ After the breakdown, the scalar field acquires a non null
vacuum expectation value (v.e.v.), ${\langle }\varphi {\rangle }=v,$ one
readily gets:\ ${\langle }\varphi ^{\ast }\varphi {\rangle }=v^{2}-\zeta
/2\lambda {+}\sqrt{\left( \zeta /2\lambda \right) ^{2}-\mu ^{2}/\lambda }$.
On the other hand, the condition for a minimum reads as: $\mu ^{2}+{\zeta }%
v^{2}+{\lambda }v^{4}=0.$ In the broken phase, the complex scalar field is
parametrized by $\varphi =v+H+i\theta $, where $\theta $ is the would-be
Goldstone boson and $H$ is the Higgs scalar, both with vanishing v.e.v.'s $({%
\langle }H{\rangle }={\langle }\theta {\rangle =0).}$ By replacing this
parametrization relation into the action (\ref{action1}), adopting the 't
Hooft gauge {\cite{thooftg} (}$S_{{\rm gf}}=\int d^{3}x\biggl\{-\frac{1}{%
2\xi }(\partial ^{m}A_{m}-\sqrt{2}\xi M_{A}\chi )^{2}\biggr\}${), and
finally }taking only the bilinear and Yukawa interaction terms, one has: 
\begin{eqnarray}
{S}^{{\rm SSB}} &=&\int {d^{3}x}\biggl\{-{\frac{1}{4}}F^{\mu \nu }F_{\mu \nu
}+{\frac{1}{2}}M_{A}^{2}A^{\mu }A_{\mu }+{\overline{\psi }_{+}}(i{%
\rlap{\hbox{$\mskip 1 mu /$}}\partial -m_{{\rm eff}}}){\psi }_{+}+{\overline{%
\psi }_{-}}(i{\rlap{\hbox{$\mskip 1 mu /$}}\partial +m_{{\rm eff}}}){\psi }%
_{-}+\partial ^{\mu }H\partial _{\mu }H-M_{H}^{2}H^{2}+  \label{SSSB} \\
&&+{\partial ^{\mu }}\theta {\partial _{\mu }}\theta -M_{\theta }^{2}\theta
^{2}-{\frac{1}{2\xi }}(\partial ^{\mu }A_{\mu })^{2}-2yv(\overline{\psi }%
_{+}\psi _{+}-\overline{\psi }_{-}\psi _{-})H-e_{3}\left( \overline{\psi }%
_{+}{\rlap{\hbox{$\mskip 1 mu /$}}A}\psi _{+}+\overline{\psi }_{-}{%
\rlap{\hbox{$\mskip 1 mu /$}}A}\psi _{-}\right) \biggr\},  \nonumber
\end{eqnarray}
where $\xi $ is a dimensionless gauge parameter and the mass generated by
the SSB are: $M_{A}^{2}=2v^{2}e_{3}^{2}~$(Proca mass), $M_{H}^{2}=2v^{2}(%
\zeta +2\lambda v^{2})~$(Higgs mass), $m_{{\rm eff}}=m_{e}+yv^{2}$(effective
electron mass), and $M_{\theta }^{2}=\xi M_{A}^{2}$. The latter corresponds
to non-physical poles in the gauge and $\theta $-field propagator. \ Their
effects are mutually canceled as already known from the study of the
unitarity in the 't Hooft gauge \cite{Unitarity}.

At this point, it is instructive to highlight that from now on our relevant
``physical'' action is the one obtained in the broken phase, that is, $%
S^{SSB}$. Based on this broken action, one derives the essential results of
this paper. In this tree-level broken action, no register of $V,$ the scalar
potential, remains. In fact, we can assert that the main role of the
sixth-order potential is basically to determine the occurrence of the
spontaneous symmetry breaking in a planar theory, where one uses to require
renormalisation. On physical grounds, one can say that the sixth-power
potential can be replaced by a fourth-power form $\left( \lambda \varphi ^{4}%
\text{-type}\right) $, also able to induce the SSB and the phase transition,
without effectively changing the features of the broken phase. The reason to
use a sixth-power scalar self-interacting potential is exclusively that it
is the most general renormalizable potential in Planar Quantum
Electrodynamics (a quantum field theory reason). Written at tree-level, the
action (\ref{SSSB}) exhibits no vertex with more than three legs, which in
turn contribute to loop diagrams. So, at this level of approximation, loop
contributions are excluded from the physical model. One knows these loop
diagrams correct the classical theory, but in a pertubation perspective,
their contribution may not be enough to modify the essence of a result
constructed at tree-level. This fact justifies the omission of the scalar
interaction terms of higher order ($H^{2},H^{3},H^{4},$...) at this level.
Although, these higher-order terms need to be considered when one desires to
analyze radiative corrections or one tries to pass from this microscopic
theory to a phenomenological model where the fermions and gauge fields are
integrated out.

In the low-energy limit (Born approximation), the two-particle interaction
potential is given by the Fourier transform of the two-particle scattering
amplitude \cite{Sakurai}. \ It is important to stress that in the case of
the non-relativistic M\"{o}ller scattering, one should consider only the
t-channel (direct scattering) \cite{Sakurai} even for distinguishable
electrons, since in this limit they recover the classical notion of
trajectory. From the action (\ref{action1}), there follow the Feynman rules
for the interaction vertices: $V_{\pm H\pm }=\pm 2ivy;V_{\psi A\psi
}=ie_{3}\gamma ^{\mu }$, so that the $e^{-}e^{-}$ scattering are written as
bellow: 
\begin{eqnarray}
-i{\cal M}_{\pm H\pm } &=&\overline{u}_{\pm }(p_{1}^{^{\prime }})(\pm
2ivy)u_{\pm }(p_{1})\left[ \langle HH\rangle \right] \overline{u}_{\pm
}(p_{2}^{^{\prime }})(\pm 2ivy)u_{\pm }(p_{2});  \label{A1} \\
-i{\cal M}_{\pm H\mp } &=&\overline{u}_{\pm }(p_{1}^{^{\prime }})(\pm
2ivy)u_{\pm }(p_{1})\left[ \langle HH\rangle \right] \overline{u}_{\mp
}(p_{2}^{^{\prime }})(\mp 2ivy)u_{\mp }(p_{2});  \label{A2} \\
-i{\cal M}_{\pm A\pm } &=&\overline{u}_{\pm }(p_{1}^{^{\prime
}})(ie_{3}\gamma ^{\mu })u_{\pm }(p_{1})\left[ \langle A_{\mu }A_{\nu
}\rangle \right] \overline{u}_{\pm }(p_{2}^{^{\prime }})(ie_{3}\gamma ^{\nu
})u_{\pm }(p_{2});  \label{A3} \\
-i{\cal M}_{\pm A\mp } &=&\overline{u}_{\pm }(p_{1}^{^{\prime
}})(ie_{3}\gamma ^{\mu })u_{\pm }(p_{1})\left[ \langle A_{\mu }A_{\nu
}\rangle \right] \overline{u}_{\mp }(p_{2}^{^{\prime }})(ie_{3}\gamma ^{\nu
})u_{\mp }(p_{2});  \label{A4}
\end{eqnarray}
where $\langle HH\rangle $ and $\langle A_{\mu }A_{\nu }\rangle $ are the
Higgs and massive photon propagators. Expressions (\ref{A1}) and (\ref{A2})
represent the scattering amplitudes for electrons of equal and opposite
polarizations mediated by the Higgs particle, whereas Eqs. (\ref{A3}) and (%
\ref{A4}) correspond to the massive photon as mediator.

The spinors $u_{\pm }(p)$ stand for the positive-energy solution of the
Dirac equation $\left( \rlap{\hbox{$\mskip1 mu /$}}p\mp m\right) u_{\pm
}(p)=0$. We adopt the following conventions $\eta _{\mu \nu }=(+,-,-),$ $%
\left[ \gamma ^{\mu },\gamma ^{\nu }\right] =2i\epsilon ^{\mu \nu \alpha
}\gamma _{\alpha }$, $\gamma ^{\mu }=(\sigma _{z},-i\sigma _{x},i\sigma
_{y}),$ whence one obtains

\begin{equation}
u_{+}(p)=\frac{1}{\sqrt{N}}\left[ 
\begin{array}{c}
E+m \\ 
-ip_{x}-p_{y}
\end{array}
\right] ,\text{ \ \ }u_{-}(p)=\frac{1}{\sqrt{N}}\left[ 
\begin{array}{c}
ip_{x}-p_{y} \\ 
E+m
\end{array}
\right] ,
\end{equation}
with $N=2m(E+m)$ being the normalization constant that assures $\overline{u}%
_{\pm }(p)u_{\pm }(p)=\pm 1$. Working in the center-of-mass frame \cite
{Int-Journal}, \cite{Tese}, the scattering amplitudes ${\cal M}%
_{higgs}=-2v^{2}y^{2}\left( \overrightarrow{k}^{2}+M_{H}^{2}\right) ^{-1},$ $%
{\cal M}_{gauge}=+e_{3}^{2}\left( \overrightarrow{k}^{2}+M_{A}^{2}\right)
^{-1}$ reveal to be independent of the spin polarization. Evaluating now the
Fourier transform of the total amplitude scattering (${\cal M}_{total}={\cal %
M}_{higgs}+{\cal M}_{gauge}$), the following interaction potential comes
out: 
\begin{equation}
V^{CM}(r)=-\frac{1}{2\pi }\biggl[2v^{2}y^{2}K_{o}(M_{{\small H}%
}r)-e_{3}^{2}K_{o}(M_{A}r)\biggr].  \label{PotencialQED1}
\end{equation}
Considering equal Higgs and Proca masses $\left(
M_{H}=M_{A}\Longleftrightarrow e_{3}^{2}=\zeta +2\lambda v^{2}\right) $, the
potential (\ref{PotencialQED1}) takes the form 
\begin{equation}
V^{CM}(r)=CK_{o}(M_{A}r),\text{ with: \ }C=-\frac{1}{2\pi }\biggl[%
2v^{2}y^{2}-e_{3}^{2}\biggr].  \label{PotentialQED2}
\end{equation}
It becomes attractive whenever $C<0$, that is, $2v^{2}y^{2}>e_{3}^{2}$. Now,
it is necessary to point out that this result is independent of the
spin-polarization of the scattered electrons, whereas the potential enclosed
in Ref. \cite{N.Cimento}, derived at the same physical conditions, shows an
erroneously inversion of sign (for the gauge interaction between
antiparallel-spin electrons). A spin dependence will arise in the low-energy 
$e^{-}e^{-}$ potential only when one considers the \ presence of the
Chern-Simons term, resulting in a Maxwell-Chern-Simons-Proca model \cite
{MCS2}.

\section{Wavefunction properties and Numerical Analysis}

Having determined the interaction potential, one must now look for the\
numerical evaluation of the binding energy associated to the $e^{-}e^{-}$
pairs. In the non-relativistic limit, the complete two-dimensional
Schr\"{o}dinger equation (supplemented by the Bessel-$K_{0}$ potential) 
\begin{equation}
\frac{\partial ^{2}\varphi (r)}{\partial r^{2}}+\frac{1}{r}\frac{\partial
\varphi (r)}{\partial r}-\frac{l^{2}}{r^{2}}\varphi (r)+2\mu _{{\rm eff}%
}[E-CK_{o}(M_{A}r)]\varphi (r)=0~  \label{diff1}
\end{equation}
yields the energy of the two interacting particles. Here, $\mu _{{\rm eff}}=%
\frac{1}{2}m_{{\rm eff}}$ $\varphi (r)$ is the effective reduced mass of the 
$e^{-}e^{-}$ system and $\varphi $ represents the (relative) spatial part of
the complete antisymmetric 2-electron wavefunction:$\;\Psi
(r_{1},s_{1,}r_{2},s_{2})=\psi ({\bf R})\varphi ({\bf r})\chi \left(
s_{1},s_{2}\right) ,~$while $\psi ({\bf R})$, $\chi \left(
s_{1},s_{2}\right) $ stand for the center-of-mass and the spin wave
functions.

For a numerical solution of the Schr\"{o}dinger equation, we employ the
variational method. In this respect, we take as starting point the choice of
a wave function that stands for the generic features of the $e^{-}e^{-}$
state: the trial function, whose definition must observe some conditions,
such as the asymptotic behavior at infinity, the analysis of its free
version and its behavior at the origin. With the help of the transformation $%
\varphi (r)=\frac{1}{\sqrt{r}}~g(r)$, Eq.(\ref{diff1}) is transformed into 
\begin{equation}
\frac{\partial ^{2}g(r)}{\partial r^{2}}-\frac{l^{2}-\frac{1}{4}}{r^{2}}%
g(r)+2\mu _{{\rm eff}}[E-CK_{o}(M_{A}r)]g(r)=0,
\end{equation}
whose free version $\left( V(r)=0\right) $ for zero angular momentum ($l=0$)
state simplifies to 
\begin{equation}
\left[ \frac{\partial ^{2}}{\partial r^{2}}+\frac{1}{4r^{2}}+k^{2}\right]
u(r)=0~.  \label{diff4}
\end{equation}
Its general solution is given by $u(r)=B_{1}\sqrt{r}J_{0}(kr)+B_{2}\sqrt{r}%
Y_{0}(kr)$, with $B_{1}$ and $B_{2}$ being arbitrary constants and $k=\sqrt{%
2\mu _{{\rm eff}}E}$. In the $r\rightarrow 0$ limit, the solution to Eq.(\ref
{diff4}) approaches $u(r)\longrightarrow \sqrt{r}+\lambda \sqrt{r}\ln r.$
Since the second term in Eq.(\ref{diff4}) behaves like an attractive
potential, $-1/4r^{2}$, this implies the unphysical possibility of obtaining
a bound state ($E<0$) even for $V(r)=0$ \cite{Chadan}. Among the infinite
number of self-adjoint extensions of the differential operator $%
-d^{2}/dr^{2}-1/4r^{2}$, the only physical choice corresponds to the
Friedrichs extension ($B_{2}=0$), which behaves like $\sqrt{r}$ at the
origin, indicating this same behavior for $u(r)$. The complete equation, $%
V(r)\neq 0$, will preserve the self-adjointness of free Hamiltonian, if the
potential is ``weak'' in the sense of the Kato condition: $\int_{0}^{\infty
}r(1+|\ln (r)|)|V(r)|dr<\infty .$ This condition also sets up a finite
number of bound states (discrete spectrum) and the semi-boundness of the
complete Hamiltonian. Provided that the Bessel-$K_{0}$ potential, given by
Eq. (\ref{PotentialQED2}), satisfies the Kato condition, the
self-adjointness of the total Hamiltonian is assured and the existence of
bound states is allowed. On the other hand, at infinity, the trial function
must vanish asymptotically in order to fulfill square integrability.
Therefore, a good and suitable trial function (for $l=0$) could be taken by 
\begin{equation}
g(r)=\sqrt{r}\exp (-\beta r)~,  \label{funcaoteste1}
\end{equation}
where $\beta $ is a free spanning parameter to be numerically fixed in order
to minimize the binding energy.

Once the trial function is already known, it still lacks a discussion on the
physical parameters $(\nu ^{2},e_{3}^{2},y^{2})$ that compose the
proportionality constant, $C$, of the Bessel potential, in such a way that
numerical values may be attributed to them. The vacuum expectation value, $%
v^{2},$ indicates the energy scale of the spontaneous breakdown of the $%
U(1)- $local symmetry. This is a free parameter, being usually determined by
some experimental data associated to the phenomenology of the model under
investigation, as occurs in the electroweak Weinberg-Salam model, for
example. On the other hand, the $y$ parameter measures the coupling between
the fermions and the Higgs scalar, working in fact as an effective constant
that embodies contributions\ of all possible mechanisms of electronic
interaction via Higgs-type (scalar) excitations, as the spinless bosonic
interaction mechanisms: phonons, plasmons, and other collective excitations.
This theoretical similarity suggests an identification of the field theory
parameter with an effective electron-scalar coupling (instead of an
electron-phonon one): $y\rightarrow \lambda _{{\rm es}}$. Specifically, in
QED$_{3}$, the electromagnetic coupling constant squared, $e_{3}^{2}$, has
dimension of mass, rather than the dimensionless character of the usual
four-dimensional QED$_{4}$ coupling constant. This fact might be understood
as a memory of the third dimension that appears (into the coupling constant)
when one tries to work with a theory intrinsically defined in three
space-time dimensions. This dimensional peculiarity could be better
implemented through the definition of a new coupling constant in three
space-time dimensions \cite{Kogan},\cite{Randjbar}: $e\rightarrow e_{3}=e/%
\sqrt{l_{\perp }}$, where $l_{\perp }$ represents a length orthogonal to the
planar dimension. The smaller is $l_{\perp }$, smaller is the remnant of the
frozen dimension, larger is the planar character of the model and the
coupling constant $e_{3}$, what reveals its effective nature. In this sense,
it is instructive to notice that the effective value of $e_{3}^{2}$\ is
always larger than $e^{2}=1/137$\ whenever $l_{\perp }$\ $<1973.23$\ \AA ,
since 1 (\AA )$^{-1}=1973.26$\ $eV$. This particularity broadens the
repulsive interaction for small $l_{\perp }$ and requires an even stronger
Higgs contribution to account for a total attractive interaction.

The following Table, constructed for zero angular momentum state $\left(
l=0\right) $, has as input data the three parameters $(\nu ^{2},l_{\perp
},y),$ while the output parameters are: $\beta -$ the minimization
parameter, $E_{e^{-}e^{-}}-$ the $e^{-}e^{-}$ binding energy, and $\langle
r\rangle -$ the average-length of the wavefunction$.$ 
\begin{table}[h]
\begin{tabular}{|l|l|l|l|l|l|l|l|}
\hline
$v^{2}(meV)$ & $l_{\perp }(\text{\AA })$ & $y$ & $C_{s}$ $(meV)$ & $M_{H}$ $%
(meV)$ & $\beta $ & $E_{e^{-}e^{-}}$ $(meV)$ & $\langle r\rangle $ $\left( 
\text{\AA }\right) $ \\ \hline
120.0 & 15.0 & 2.1 & -15.7 & 480.0 & 63.1 & -74.7 & 15.6 \\ \hline
120.0 & 14.0 & 2.1 & -4.8 & 496.8 & 35.8 & -19.7 & 27.6 \\ \hline
120.0 & 13.0 & 2.2 & -8.6 & 515.6 & 47.2 & -37.8 & 20.9 \\ \hline
100.0 & 12.0 & 2.6 & -24.2 & 489.9 & 81.1 & -120.2 & 12.2 \\ \hline
100.0 & 12.0 & 2.5 & -8.0 & 489.9 & 45.9 & -35.2 & 21.5 \\ \hline
100.0 & 10.0 & 2.7 & -2.9 & 536.6 & 27.8 & -11.0 & 35.5 \\ \hline
100.0 & 10.0 & 2.8 & -20.4 & 536.6 & 72.1 & -97.6 & 13.7 \\ \hline
100.0 & 6.0 & 3.5 & -8.0 & 692.8 & 45.9 & -32.5 & 21.5 \\ \hline
80.0 & 6.0 & 3.9 & -5.4 & 619.6 & 37.6 & -21.4 & 26.2 \\ \hline
70.0 & 4.0 & 5.1 & -6.6 & 709.9 & 41.7 & -26.2 & 23.6 \\ \hline
60.0 & 8.0 & 3.9 & -4.0 & 464.7 & 33.1 & -16.7 & 29.8 \\ \hline
\end{tabular}
\caption{Input data ($v^{2},l_{\perp },y$) and output data ($E_{e-e},\langle
r\rangle $) for the Schr\"{o}dinger Equation}
\label{table1}
\end{table}

\section{Final Remarks}

The numerical data of Table \ref{table1} show that the attractive Bessel
potential, derived for a non-relativistic regime, may effectively promote
the formation of $e^{-}e^{-}$ bound states. The procedure here carried out
puts in evidence that, by properly fitting the free parameters of the model,
one can obtain bound states of the order of $10-100$ $meV$ and wavefunction
average-length in the range $10-30$ \AA , which may reveal the suitability
of the framework here adopted to address the issue of electron-electron
condensation in the realm of parity-preserving planar systems. Finally, we
can assert that the photon Proca mass, generated by the SSB, plays the same
role of the topological mass $\left( \vartheta \right) $ in that it
determines the Coulomb interaction screening and the Meissner effect,
without breaking parity-symmetry. The data exhibited in Table \ref{table1}
concern an s-wave state: $l=0$ and spin singlet ($\uparrow \downarrow ,S=0$%
). According to the results of this paper, we conclude by stressing the
fundamental role played by the Higgs mechanism in QED$_{3}$ as essential for
the appearance of an attractive $e^{-}e^{-}$ potential.

Final comments on the general procedure here employed are still necessary.
We conceive this paper as the first part of a two-stage project, described
as follows. At the first stage, we have a microscopic model, where the
presence of all degrees of freedom (relative to fermions, vector and scalar
bosons) is necessary. In this moment, our purpose is to exhibit a
microscopic mechanism, at tree level, able to yield the electronic pairing.
Since we are bound to the tree-approximation, higher powers in $H$ need not
be considered whenever computing the transition amplitude from which we read
off the inter-particle potential. This is exactly what we have done here. In
a second stage of development, still to be performed, one should take into
account the high-order terms in $H$ (stemming from the Higgs potential and
from the electronic coupling). A functional integration on the fermions and
vector bosons must be carried out, yielding an effective functional with
dependence only on the Higgs-scalar field. We think this functional may
exhibit a Ginzburg-Landau-like form, if the sixth-order potential is
replaced by a quartic-order potential. To our mind, the Higgs field, which
represents the scalar excitations (relative to the vacuum expectation
value), will play (after the functional integration) the same role as the
order parameter plays in the G-L model: that of a fluctuation field in the
context of a mean-field theory.

J.A. Helayel-Neto expresses his gratitude to CNPq for the financial help.

\end{document}